# Integrated approaches in physics education: A graduate level course in physics, pedagogy, and education research

Michael C. Wittmann and John R. Thompson
Department of Physics and Astronomy
College of Education and Human Development
Center for Science and Mathematics Education Research
University of Maine
Orono ME 04469-5709
wittmann@umit.maine.edu

## Abstract

We describe a course designed to help future educators build an integrated understanding of the different elements of physics education research (PER), including: research into student learning, content knowledge from the perspective of how it is learned, and reform-based curricula together with evidence of their effectiveness. Course elements include equal parts of studying physics through proven curricula and discussion of research results in the context of the PER literature. We provide examples of the course content and structure as well as representative examples of student learning in the class.

## 1. INTRODUCTION

With the growth of physics education research (PER) as a research field [1,2] and the ongoing desire to improve teaching of introductory physics courses using reform-based approaches [3], there has been an opportunity to move beyond an apprenticeship model of learning about PER toward a course-driven structure. At the University of Maine, as part of our Master of Science in Teaching program, we have developed and taught two courses in *Integrated Approaches in Physics Education*. These are designed to teach physics content, PER methods, and results of investigations into student learning. Course materials were inspired by conversations in 1999 and 2000 with Noah Finkelstein (now at University of Colorado in Boulder) as described in the companion paper in this issue.[REF to be determined in publication] Materials development was led by Michael Wittmann, with assistance from Dewey Dykstra (Boise State University), Nicole Gillespie (now at the Knowles Science Teaching Foundation), Rachel Scherr (University of Maryland), and John Thompson, who later joined the University of Maine and has since modified the materials while teaching the courses. Constraints, described below, led to a pair of courses very different from those originally discussed by Finkelstein and Wittmann, but the general purpose has remained the same: to create an environment in which research on education and teaching were both promoted and fed off each other.

The goal of our courses is to build a research-based foundation for future teachers (at the high school and university level) as they move into teaching. We describe the origins of the courses and the activities that make up a typical learning cycle. We also describe one instructional unit from one course in some detail. Examples of student learning in the course provide insight into the types of reasoning our future teachers are capable of and how they use research results to guide their reasoning. (We are engaged in a large study to examine student learning of physics, pedagogy, and PER results, and plan to report extensively on these results in the future.)

Note that in this paper, we refer to "graduate students" taking the courses as GSs (though not all class participants are graduate students) and we refer to students learning physics, typically the subjects of physics education research, simply as students.

## 2.  COURSE GOALS

Our objectives in designing the *Integrated Approaches* courses are that practicing and future teachers will:
- learn relevant physics content knowledge at an appropriately deep level,
- become familiar with "best practices" research-based instructional materials, and
- gain insight into how students think about physics through education research into student learning and curriculum effectiveness.

These objectives are consistent with those of the Master of Science in Teaching (MST) program sponsored by the University of Maine Center for Science and Mathematics Education Research.  We wish for participants (GSs) to learn content in courses taught using research-guided pedagogy and curricula, including hands-on, inquiry-based methods.  We offer courses that integrate content and methods learning.  By taking such courses, students learn how to design and conduct science and math education research and are better able to interpret the results of this kind of research to benefit their target population.  They apply these ideas when carrying out their own discipline-specific education research projects as part of their master's thesis work.

The courses exist under several constraints due to the GS population targeted for the MST program.  We have designed the courses to be relevant to in-service physics teachers wanting either a deeper understanding of the physics content they are teaching, experience and exposure to physics education research, or research-based pedagogical tools.  Many from this population are teaching "out of field," and have little physics background.  Many of our MST physics students are transitioning from careers in science or engineering into careers in education, and have little pedagogical content knowledge (which we use to mean knowledge about how to represent the content appropriate to teaching) [4].  However, the courses are also taken by second- or third-year physics graduate students who are doing PER for their Ph.D. work or wishing to improve their teaching skills as they prepare for careers in academia.  This population typically has not taught outside of teaching assistantships in college courses.  Finally, we have many MST students from other science and mathematics fields.  As a result, there is a great variety in both physics and pedagogical content knowledge among the GSs.  The differences in these populations have led to interesting discussions which illustrate the importance of both physics and pedagogical content knowledge for a complete understanding of PER results and implementations, as well as a deeper understanding of student learning in physics.

## 3.  COURSE STRUCTURE

The *Integrated Approaches* courses are 3-credit graduate courses that meet twice a week for a total of 150 minutes.  We teach content knowledge, education research results, and research methods using a three-tiered structure.  Class time is spent approximately equally on each of the three elements of the courses.  A research and development project is carried out in parallel, primarily outside of class time.

## A. Course Design

We split each course into content-based units in which we discuss leading curricula, the research literature related to that material, and emphasize one or two education research methods. The instructional units for the individual courses are presented in tables 1 and 2. In addition to the primary curricula listed in the tables, we also discuss curricula and instructional strategies such as *Just-in-Time Teaching* [22] and *Physlets* [23]. The two courses are designed to be independent of each other.

| Physics content | Curriculum emphasized | Research method |
|---|---|---|
| Electric circuits | *Tutorials in Introductory Physics*[5] and materials from Gutwill et al.[6] | Analysis of free response pre- and post-test responses[7,8] |
| Kinematics | *Activity-Based Tutorials*[9,10], *RealTime Physics*[11], and *Powerful Ideas in Physical Science*[12] | Free response questions, multiple-choice surveys (Test for Understanding Graphs in Kinematics (TUG-K)[13] and Force and Motion Conceptual Evaluation (FMCE)[16]) |
| Forces and Newton's Laws | *Tutorials in Introductory Physics*[5] and *UMaryland "epistemological tutorials"*[14] | Multiple-choice surveys (Force Concept Inventory (FCI)[15] and FMCE[16]) |

TABLE 1: First semester instructional units.

| Physics content | Curriculum emphasized | Research method |
|---|---|---|
| Wave physics and sound | *Activity-Based Tutorials* [9,10] and *Physics by Inquiry* (in development) | Student interviews [17], comparing multiple-choice to free response questions [18] |
| Work-energy and impulse-momentum | *Tutorials in Introductory Physics* [5] | Student interviews [19], comparing multiple-choice to free response questions [20] |
| Heat and temperature | *UC Berkeley lab-tutorials* and *Physics by Inquiry* [21] | Classroom interactions, research-based curriculum development and modification |

TABLE 2: Second semester instructional units.

## B. A cycle of integrated instruction

To date, a typical cycle of instruction consists of:
- pretests on the physics which will be studied, to explore the depth of understanding of our GSs (many are weak in physics, and we need to know how best to help them);
- pretests on what (introductory) students might believe about this physics, to see how good a picture the GSs have of student reasoning about the topic;
- instruction on the physics using published, research-based curricula, as listed above;

- discussion of the research literature on the physics topic, typically based on papers directly related to the instructional materials, but often set up to complement and create discussion;
- homework dealing primarily with the physics and not the pedagogy; and
- a post-test on all three areas of physics, pedagogy, and research and how they intersect.

We explain the various elements of our instructional cycle before giving two examples from instruction on force and motion and on electric circuits.

Having GSs work through conceptually-oriented research-based materials is a necessary component of many teaching assistant preparation seminars. By working through instructional materials, GSs focus on conceptual understanding by building simple models of physical phenomena and looking to understand the physics that is taught in a new way. In the process, GSs with weak physics strengthen their content knowledge, while those who are stronger often see the physics from a new point of view. Our course benefits the GSs by having them work through multiple instructional materials and subsequently participate in classroom discussions comparing the pros and cons of different curricula.

These discussions can be very helpful in teaching physics content *and* pedagogical content knowledge. Which of the instructional materials is best for students might depend on the course setting, student demographics, mathematical background of the population, and so on. Debating these issues with our GSs lets them understand the role of content knowledge, pedagogical choices teachers make, and the standards of evidence used to make choices. The combination gives them a fuller picture of research-based instruction.

Curriculum discussions are guided in large part by education research results on a given topic. Students read papers on student learning of a given physics topic, evaluation of a given curriculum (in best cases, the one we are using to teach content knowledge at the time), and ways in which different models of student reasoning affect curriculum design by researchers and developers. Because we choose papers directly connected to the curricula we are studying, students can gain deeper insight into the origin of the instructional materials and the specific issues that curriculum developers were hoping to address. Because developers typically use results beyond their own work, we have a rich collection of literature to reach back to. We usually assign influential and well-known papers in PER, typically found in the 1998 AJP Resource Letter in PER [24] or more recent results as outlined in the Forum *Fall 2005 Newsletter* article [2]. We also include relevant pre-prints or drafts of papers associated with ongoing research as a way of promoting the idea of PER as an active, growing, dynamic field.

Other pedagogical issues are brought up in multiple contexts. For example, the issue of teaching certain topics from a purely microscopic or macroscopic perspective, or a combination, comes up in both the electric circuits [25] and the heat and temperature units [26]. These topics appear in different semesters, which allows students taking only one semester of the course to broach the issue, and those who take both courses to consider the issue in more than one content area.

Research methods are introduced by readings from the PER literature, and students learn research skills by carrying out research projects in the course. Skills for developing research tools such as written questions, surveys, and interviews are developed during class time. Students also spend time in class and on homework practicing data analysis. We give examples below. Furthermore, we have students learn about and practice clinical interview techniques in class before doing their own interviews in their class-based research projects. In some instances, we have students analyze video of students working in a classroom situation. By studying interactions in social groups without teaching assistants, students can gain a deeper perspective on learning in all elements of a course.

A final part of the course is to pull together physics and pedagogical content knowledge, understanding of research methodologies, analysis skills, and research-based curriculum design into research projects. These research projects were originally done individually, but are now done in small groups (2-4 students) as either large, semester-long, projects or a series of smaller projects, depending on the semester. Typically, students carry out one cycle of a research and development process. Building on a literature review, students design interview protocols and conduct individual interviews on a topic, use results to develop free-response and multiple-choice surveys to get written data, and analyze data from a relevant population to gain perspective on student reasoning about a given topic. Using their results, they must design a draft set of narrowly focused learning materials that are appropriate to the data they have gathered, the literature, and what is known about learning in physics.

## 4. MEASURING LEARNING: A UNIT ON ELECTRIC CIRCUITS

We summarize the unit on electric circuits and show data on GS learning of physics and pedagogical content knowledge in the course. In the electric circuits unit, we emphasize materials from the *Tutorials in Introductory Physics* [2] while reading papers related to the creation of the curriculum materials [7,8] and developing skills in analyzing student written responses on the associated pretest questions.

Before instruction, GSs must answer the "5 bulbs" question (Figure 1) *and* predict what an "ideal incorrect student" might answer in a similar situation. We report on their understanding of the physics below. An "ideal incorrect student" response on the "5 bulbs analysis task" would match results from the research literature and be self-consistent throughout the response (though, of course, students aren't always consistent when giving wrong answers). GSs analyze typical responses by categorizing 20 anonymous student responses to the "5 bulbs" question [7,8] – before reading the research results on this question. One class period is spent on discussions of different categorizations. We report on results below. Students work through curriculum pieces that begin with simple series and parallel circuits and progress through RC circuits. Students consider other curricula for teaching current (listed above) and discuss the merits and weaknesses of each. Finally, they are tested on their understanding both of the physics and the research and instruction choices that determine one's lessons. To show understanding, they must refer to the correct physics and the literature on student learning, as is appropriate. Tests have in-class and take-home components to allow for evaluation of more time-consuming analyses of student thinking. In sum, we teach – and test – whether students themselves learn the correct physics concepts and whether they can predict, analyze, and classify incorrect reasoning they are likely to encounter when teaching. (In later parts of the course, we also ask students to suggest, design, or critique instructional materials which address typical incorrect responses.)

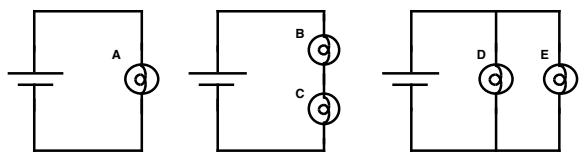

FIG. 1: "5 bulbs" question. Students must rank the brightness of each bulb.
Correct response for ideal batteries and bulbs: A = D = E > B = C.

## A. Content knowledge: Circuits.

We have several years of GS data for the 5–bulbs pretest question. This data was obtained *before* GSs read the relevant literature on the research [7,8]. We have broken down the results two different ways. First, the obvious correct/incorrect distribution: thirteen of the 23 Graduate Students (57%) provided the correct ranking of the bulbs (A=D=E>B=C), with correct reasoning, and with varying levels of detail. A second breakdown is by physics/non-physics backgrounds: Of those with physics backgrounds, 12 of 15 were correct (80%); of those with non-physics backgrounds, 1 of 8 was correct (13%). This question was given in a similar course at the University of Maryland, and 21 of 26 (81%) physics graduate students gave correct responses. In comparison, data from UW [7,8] show that, typically, only 15% of students answer the question correctly, even after instruction. Our non-physics GSs are typical of a larger population of students. Clearly non-physics GSs who may very well teach this simple physics topic in the future need help learning the physics. We can also not assume that all physics GSs know the material. Common incorrect responses are consistent with introductory physics student answers (including "current used up" and "battery provides constant current" answers). On post-tests, GSs typically perform at the 100% level for the entire population on certain content-based questions that are more difficult than the 5-bulbs question. See figure 2 for an example of a post-test question [27].

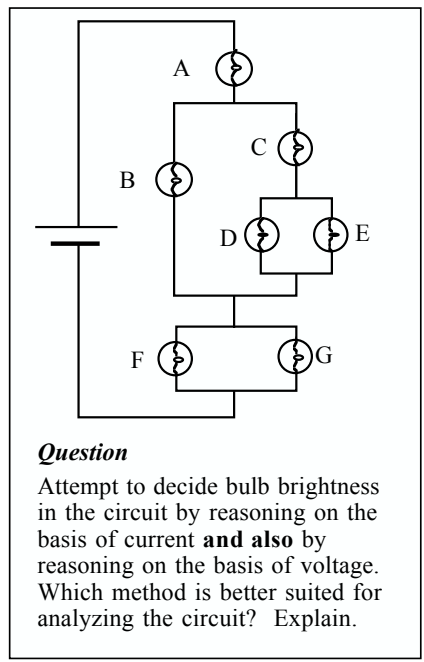

FIG 2. Representative post-test question for the Electric Circuits unit administered to GSs.

## B. Example research method: Analyzing pretests.

We have had 23 GSs do the "5 bulbs analysis task," described above. They have used at least 10 different analysis methods. What fascinates us is the natural language they bring to the problem before receiving training in analysis and before discussing analysis methods with each other. Broad categories include whether students' wrong answers are measured, or whether the analyzer attends to what students are doing well. Most GSs focus on incorrect answers, as they have been implicitly trained to do through their own academic careers. Few emphasize valuable elements of student responses. More subtle issues include whether GSs organize the analysis along individual responses (of which there are many, and nearly no repeats) or try to encode likely reasoning models into the models. The former is ungainly, while the latter incorporates often unquestioned theoretical orientations and assumptions at the first step of analysis. Finally, we observe that our students typically only use one analysis method, rather than considering a single data from multiple perspectives. Many unspoken assumptions must be stated and questioned in the ensuing class discussion of the analysis. In summary, our GSs are novices at analysis – not naïve, but definitely inexperienced. The *Integrated Approaches* courses are designed to help GSs understand their unquestioned assumptions about students, why they value correct elements of reasoning so little, and how they assume that students construct their knowledge of physics.

## C. Understanding pedagogy: Analyzing the intent of questions.

In one electric circuits post-test question, the GSs were given 5 students' responses to an expanded 5-bulbs question. The first two parts of the modified question were the same as in figure 1. A third question was added to this pretest, based on a personal communication from Bradley S. Ambrose at Grand Valley State University. It asked the students to rank the *currents through the battery* for the three circuits in the figure. It should be noted that the GSs had never seen the third pretest question before this point in the semester, though the general issue had been discussed in class. The GSs were asked to analyze the student ranking and reasoning for each question independently; they were then asked to discuss the purpose of the third question – what insight could it give into students' reasoning? We have anonymous data from the three question analysis asked on take-home exams using this modifications. (GSs also had to analyze student responses to each question and discuss consistency of student responses as part of the take-home test.)

GS responses illustrate the types of learning we wish them to attain. A biology student with little background in physics stated:

> [The current question] gives insight into whether or not the students truly consider the battery as a constant current source. The correct ranking of B and C being equal, but dimmer than A because current is "shared" might not fully bring forth the idea of the battery as a constant current source. This is shown in the answers of Student 5. ... Although Student 1 shows a similar idea in question 1 that the battery is a constant current source and doesn't state it explicitly, the answer given to question 2 confirms the model.

Note that the student compares two student responses to illustrate the value of the question in giving a more complete interpretation of student thinking. A physics student (familiar with *Tutorials* but not the unit on circuits) stated:

> [The current question] is useful in prying reasoning from the students. By asking what is happening at the battery, it is far easier to elicit a clear "constant current" model, if that is indeed a model which the student uses. It also allows us to discover if a student is thinking holistically or piece-wise, by comparing what the student believes is going on in the battery to … the rest of the circuit.

In this response, the difference between holistic or piece-wise analysis of the circuit is pointed out, as described in the literature [7, 8]. In both examples, we find that students after instruction are able to carefully interpret student reasoning in a way that is useful for interpreting curriculum materials and facilitation of student learning.

We have only discussed a small subset of the data we have, not all of which has been analyzed. In our teaching, we seek to triangulate between different data sources which include: classroom video of students throughout an entire semester; written pre- and post-tests on content understanding; written pre- and post-tests on pedagogical content, *e.g.,* GSs' responses of what they feel might be the most common incorrect responses on pretests, standardized surveys, and examination questions; and the GS mini-group project materials, including pretests, analysis, and curriculum. Data exist for all units and continue to be gathered. We will report on results in future publications.

## 5. INSTRUCTIONAL UNIT: MOTION AND FORCE

We teach motion and force after we teach electric circuits because we have found that GSs are more likely to respond to the PER content of circuits than to forces. Once they are attuned to discussing PER results, it is easier to return to earlier topics from typical introductory physics courses.

### A. Summary of activities

The unit on motion and force strongly emphasizes the research methods associated with standardized tests such as the Test for Understanding Graphs in Kinematics (TUG-K) [13], Force Concept Inventory (FCI) [15], and Force and Motion Conceptual Evaluation (FMCE) [16]. The GSs, many very familiar with the actual physics, are typically less familiar with research into student learning of the topic. The GSs read papers by Trowbridge and McDermott [28,29], Minstrell [30], Reif and Allen [31], diSessa [32], Brown and Clement [33], Dykstra [34,35], and Hammer and Elby [36]. Curricula that are studied include the *Activity-Based Tutorial* on velocity, using motion sensors [9], *Just-in-Time Teaching* [22] and *Physlets* [23], *Powerful Ideas in Physics Science* [12], *Tutorials in Introductory Physics* [5], *RealTime Physics* labs [11], as well as an example of a "refining intuitions" tutorial from Elby[36]. While studying these materials, GSs are engaged in mini-group projects in which they develop free-response and multiple-choice questions (on the same topic), administer the questions to students, analyze the results, and compare data from the two question formats. Finally, they must answer the FCI and the FMCE to the best of their ability *and* as they would expect an ideal, incorrect student to answer. They must also justify the "ideal, incorrect" responses.

In summary, GSs in the units on motion and force study physics content pedagogies developed based on research, read the associated research literature, and develop their own research instruments. Many opportunities for growth in physics and pedagogical content knowledge are possible.

### B. The interplay of content and pedagogical content knowledge

Instruction on Newton's laws provides many examples for comparing instructional choices. We base our discussions on an understanding of physics and of pedagogy, including how a curriculum presents the physics and what population is best matched to it. We give only one example, though there are several in this unit, alone. When first presenting Newton's Second Law, *RealTime Physics (RTP)* [11] uses dynamic situations with a single horizontal force while *Tutorials in Introductory Physics (TiIP)* [5] uses static situations with many forces acting at once.

In *RTP*, students first study how a single force causes an acceleration, a concept which in itself is often difficult to understand. Students start with observations of low friction carts speeding up, relate that change in speed to the mass hanging from the pulley, account for the ratio of cart mass to acceleration by defining the force acting on the cart, and then consider competing forces (such as from two different fans on a single cart) as well as the variation of balancing forces leading to no motion.

In *TiIP*, students begin by considering many types of forces acting on a single, stationary object on which many forces are acting. They learn to distinguish between contact and non-contact forces, compare systems in horizontal and vertical systems, and generally use a static situation to derive the idea of Newton's Second Law. Student learning is very different compared to *RTP*. Students immediately think of balancing forces and must incorporate ideas related to Newton's Third Law much sooner. During the tutorial, there is no discussion of objects in motion, but horizontal and vertical cases (with multiple objects and many different kinds of forces being exerted). Later, in the homework, students must consider motion and the effects of changes in motion on each of the different types of forces.

The *RTP* and the *TiIP* approach differ in striking ways, yet both are shown to be effective in teaching Newton's Laws. We guide a discussion among the GSs on the possibly large differences in what and how students understand Newton's Laws after using each curriculum.

## C. Teaching of education research

We help our GSs learn to create, administer, and analyze results from a variety of research tools common to PER. For example, when analyzing the TUG-K [13] or the FMCE [16], students are given data tables with student responses and asked to build models of student reasoning about specific physics content. An example question from an examination is shown in Figure 3.

In other questions, we have asked students to decide why certain questions are not analyzed on the FMCE [37], though they are nearly identical to others that are analyzed. The purpose is to help the GSs practice and understand the ways in which standardized tests are useful when analyzed in part, rather than analyzed as a whole (such as with a normalized gain). Results are encouraging, in that students accurately connect student reasoning to curricula and the research literature in which results are discussed.

## 6. DISCUSSION

In this paper, we have discussed two elements of our courses in *Integrated Approaches in Physics Education*. These two examples summarize a part of the first semester content. A similar course structure is used in our second semester, while emphasizing different content and research methods. The two semesters together are designed to teach a conceptual perspective of physics content, develop the basic skills for education research methods, and bring about familiarity with the research-based curricula and the research literature that supports their effectiveness in the classroom.

Consider 9 questions taken from the Force and Motion Conceptual Evaluation (FMCE, questions 8–13 and 27–29). The table below contains data for 3 students who took the FMCE before any instruction on motion or forces.

| Question | 8 | 9 | 10 | 11 | 12 | 13 | 27 | 28 | 29 |
|---|---|---|---|---|---|---|---|---|---|
| Correct answer | a | a | a | a | a | a | a | a | a |
| Student 1 | g | d | b | g | d | b | g | d | b |
| Student 2 | f | a | c | f | d | a | g | d | b |
| Student 3 | g | d | b | g | d | a | a | d | a |

a. Analyze each of the student responses to these questions. In your answer, include the following:
   - the student view of motion and force (*e.g.* a model to describe their responses)
   - an explicit description of the consistency of their thinking
   - references, where appropriate, to the research literature (*e.g.* which papers discuss the models that you describe)
b. For each student, describe which of the curricula that we have studied would help that student develop a better understanding of the physics. To answer this question, the following comments are important:
   - choose one curriculum only, if many are possible
   - if possible, describe which specific activities in the chosen curriculum would help the student (refer to question number or a description of the activity)
   - if necessary, include curricula we have not discussed in class but have been described in the readings

FIG. 3. Example FMCE analysis question, given on the take-home portion of a midterm examination. Data are from actual student responses.

　　　The courses we describe have been firmly placed within the curriculum for the recently created Master of Science in Teaching program at the University of Maine. There is strong support for these courses within the Center and the Department of Physics and Astronomy; a similar course structure exists in a course on Earth Systems Science Education Research [38]. At the same time, the *Integrated Approaches* courses have been modified based on situational constraints. The largest change is that the courses were initially envisioned as a two-semester sequence but were altered to be two stand-alone courses due to the many interdisciplinary courses taken by MST students. The courses will continue to be modified in order to meet the needs of the students taking the courses, and to reflect the state of the field of PER. For example, in the electric circuits unit in the future, we will include a discussion of the impact of simulations in conjunction with the materials from the *Tutorials in Introductory Physics,* as reported by researchers at the University of Colorado.[39] This brings in new technologies and adds to the discussion of the macroscopic/microscopic pedagogy issue.
　　　Because we have several different student populations, it is possible to differentiate between them when considering the learning of physics and pedagogical content knowledge. We find that most physics GSs enter the course with good conceptual knowledge, but that most of the non-physics GSs in the course need help in developing their physics understanding. We find that most of the GSs are novices in analyzing education research data from written, free-response questions and survey, multiple-choice questions. We will report on these results in more detail in an upcoming paper.

## Acknowledgments

This work was supported in part by US Dept. of Education grant R125K010106. Noah Finkelstein and Ed Price gave helpful commentary on draft versions of this paper. An earlier version of this paper was published in the Spring 2006 APS Forum on Education Newsletter.